# Feel the Static and Kinetic Friction

Felix G. Hamza-Lup[1], William H. Baird[2]

[1]Computer Science and Information Technology, [2]Chemistry and Physics
Armstrong Atlantic State University, Savannah, USA
{`Felix.Hamza-Lup, William.Baird`}@armstrong.edu

**Abstract**
Multimodal simulations augment the presentation of abstract concepts facilitating theoretical models understanding and learning. Most simulations only engage two of our five senses: sight and hearing. If we employ additional sensory communication channels in simulations, we may gain a deeper understanding of illustrated concepts by increasing the communication bandwidth and providing alternative perspectives.
We implemented the sense of touch in 3D simulations to teach important concepts in introductory physics. Specifically, we developed a visuo-haptic simulation for friction. We prove that interactive 3D haptic simulations – if carefully developed and deployed – are useful in engaging students and allowing them to understand concepts faster. We hypothesize that large scale deployment of such haptic-based simulators in science laboratories is now possible due to the advancements in haptic software and hardware technology.

## 1   Introduction

Simulators are often used to illustrate abstract concepts that are generally difficult to grasp. Students may gain a deeper understanding of these concepts when using simulators that provide one or many accurate contexts for them [1]. Due to the flexibility of simulators in terms of configuration and range of options, they are sometimes superior to traditional laboratory experiments. For instance, simulators can be used to illustrate concepts that would otherwise require expensive equipment. Even in cases where the equipment itself is inexpensive, such as the wooden blocks and inclined planes commonly used in laboratory exercises studying friction, there is a limit to the number of different physical realizations of the block and board that can be either purchased or stored. Students can also manipulate components of a simulated environment in ways that are impossible in some traditional experiments. In the aforementioned case of the wooden block and inclined plane, the values of the frictional coefficients can be varied smoothly and over an arbitrary range at will within a simulator.

Furthermore, one of the problems with the physical study of friction is the relative lack of reproducibility; a student who places the block in a slightly different location on the plane (or who happens to put the block down on a different side) may get significantly different results between trials. In this case, the laboratory exercise may cause confusion rather than enhance concept understanding.

Haptics is the science of applying the tactile sense to computer applications, enabling users to receive tangible feedback, in addition to receiving other cues (e.g., auditory and/or visual). The tactile sense is frequently employed to understand the world around us [2]. With haptic devices, students are able to experience tactile sensations in the simulated environment, enabling a potentially deeper understanding of concepts and phenomena.

The paper is structured as follows. In Section 2 we present research and development work related to our visuo-haptic simulator. In Section 3 we focus on the user interaction from the visual and haptic perspective. In Section 4 we present the experimental setup deployed in a classroom environment and provide an analysis of the results. We conclude with a few remarks regarding the development of the visuo-haptic simulator and assessment in Section 5.

## 2      Related Work

Interest in the field of haptics has increased in recent years, mainly due to the potential applications in entertainment (e.g., games) and medical training. Our current focus is to develop and assess the efficiency of haptic applications in education.

There are several research programs focusing on applications of haptics into higher education. Stanford University has developed a low-cost haptic device, the haptic paddle, to augment teaching undergraduate dynamic systems courses [3]. The system was adopted and modified by Rice University researchers to fit their undergraduate course needs [4]. A group from Ohio University has developed several haptics-based activities to demonstrate concepts from physics to undergraduate engineering students [5]. At the University of Michigan, two haptics interfaces, the iTouch Motor and the Box, were designed for use in a system dynamics course and an embedded control systems course [6].

Haptics use has also expanded into K-12 education. For example, an atomic force microscope allows middle and high school students to physically manipulate live viruses over the Internet, enhancing their understanding of virus morphology [7] and significantly increasing their interest in science. Haptic Virtual Manipulatives [8] have been developed to help teach mathematics to students with learning disabilities. The group at the Ohio University pushed haptics even further by developing a set of downloadable tutorials for high school physics students [9].

An interesting haptics-based system for modeling complex molecular structures has been developed, which allows students to study molecules that are too difficult to represent in a textbook using the traditional ball-and-stick method [10]. Users are able to feel forces at the molecular level using the Interactive Molecular Dynamics system by manipulating molecules in a haptic simulation [11]. Johns Hopkins University has promoted the incorporation of haptics into all levels of education. For instance, they suggested the installation of haptic interfaces in museums to help demonstrate scientific and mathematical phenomena [12]. The University of Patras in Greece developed simulators to provide instruction to children in various areas of science, including space exploration and Newton's laws [13]. What all these simulations have in common is a framework of forces that can be simulated using haptics to emphasize and enhance abstract concept understanding. In the following section we draw the spotlight on the static and kinetic friction model.

## 3    Simulating Friction

When developing visuo-haptic simulators, we look for concepts that involve forces, so we can present these concepts from a novel perspective. We chose friction since we observed that students have difficulty applying the theoretical concepts to problems. To provide a different perspective on the forces that act on a block on an inclined plane, we developed a 3D visuo-haptic simulator.

The theoretical framework defines three types of friction forces: *static*, which prevents the initial movement of an object along a surface; *kinetic*, which replaces static friction once the object is in motion; and *rolling*, which acts on a rolling object.

Static friction is defined by the inequality $F_s \leq \mu_s N$, where $F_s$ is the force of static friction, $\mu_s$ is the coefficient of static friction, and $N$ is the normal force. The maximum value of the static friction $F_s^{max}$ is equal to $\mu_s N$. Fig. 1 illustrates the forces that act on an object being pushed up an inclined plane. We can visualize the normal force $N$ (vector pointing up perpendicular to the plane), the user-applied force $F$ (pointing right), static friction $F_s$ (pointing in the opposite direction of $F$ in this case), and the force of gravity $G$.

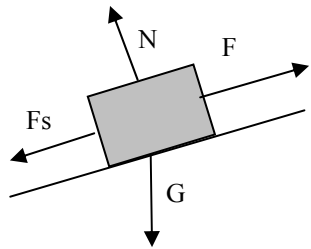

**Fig. 1.** Forces acting on a block pushed up on an inclined plane.

The fact that the static frictional force is described by an *in*equality is the ultimate source of difficulty for many students. Since all forces they have seen before are described by ordinary equalities, they tend to set $F_s = F_s^{max}$. Depending on the problem, students may not be able to realize their mistake (e.g., when an object is pushed with a force greater than $F_s^{max}$). If the force applied to an object is less than $F_s^{max}$, the use of the incorrect equality $F_s = F_s^{max}$ yields the nonphysical result that an object will move in the opposite direction of the force being applied. This dynamic component necessary to understand this phenomenon, however, cannot be illustrated in a textbook. Assume a student is given a problem like the one illustrated in Fig. 2, where the she must determine if and which way the blocks will move, and with what acceleration. Because – for the right values of each mass, angle of inclination, and coefficient of static friction – the system can move in either direction or be in equilibrium, looking only in a textbook figure, there is nothing to help the student realize that she is making an error by setting Fs = Fs$_{max}$. An interactive haptic simulation where she could feel and see the forces, however, could complement the in-classroom teaching material. This would be especially beneficial while the concepts are fresh in memory, before experimenting in the laboratory.

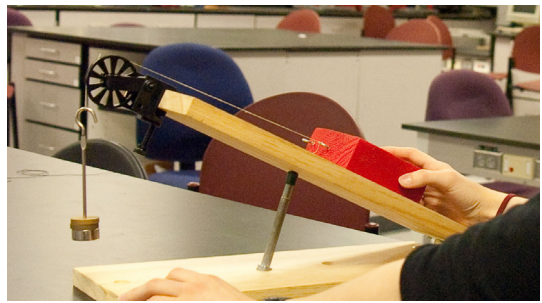

**Fig. 2.** A typical friction problem – *traditional* laboratory experiments.

In the following sections provide a description of the visual component (the graphical user interface) and the haptic component (the haptic user interface – HUI). The complexity of the system comes from the requirement to obtain an ideal perceptual integration of the visual and haptic cues while maintaining high levels of interactivity.

### 3.1 The Visual Component

We employed the H3D API [14], Extensible 3D (X3D) [15], and the Python scripting language [16] to develop the simulator.

As illustrated in Fig. 3, the visual component of the simulator consists of an inclined plane, a set of floating menus for the configuration of the experiment, and the visual pointer of the haptic device (shown as a small dot).

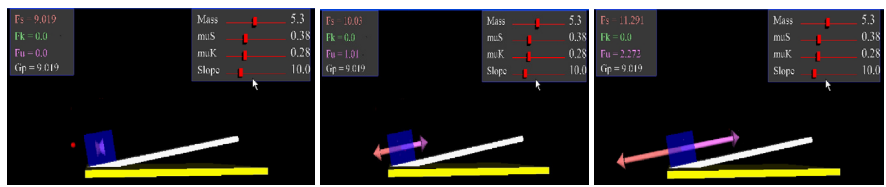

**Fig. 3.** Force magnitudes represented as arrows, as the user pushes the block up. The dot on the leftmost image near the block represents the position of the haptic pointer.

By using the menus in the heads-up display, the students can control parameters such as the block's mass, the coefficients of friction, and the slope of the plane. Such configuration changes allow students to *see* and *feel* the effects each parameter has on the forces. The magnitude of the force vectors are displayed in the other menu, enabling students to observe how these forces vary in response to configuration changes. Furthermore, the force vectors are displayed dynamically as small arrows of varying length during the interaction with the block. To obtain a different perspective of the scene, the student may change the viewpoint by rotating a disk at the bottom of the screen. The interaction can be recorded in a sequence of screenshots or small movies and used later to complement course material or laboratory sessions.

### 3.2 The Haptic Component

The HUI relies on the Novint hardware. We employed the Novint Falcon haptic device [17] because of its compatibility with the H3D API, its haptic resolution characteristics, and its affordability. The cost becomes an important aspect, as we intend to equip classrooms of thirty to forty students with one haptic interface per computer. Most physics laboratories can also be enhanced by connecting these devices to available computers (using a plug-n-play USB connection). Students can now use the Novint Falcon to interact with the virtual block and plane, and feel the resulting forces, as illustrated in Fig. 4.

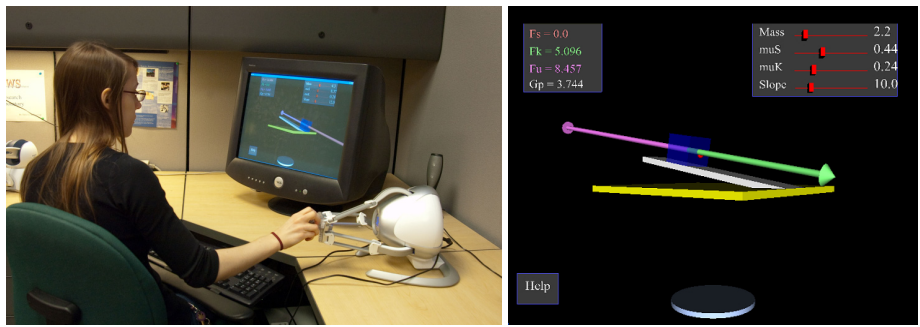

**Fig. 4.** Student using the haptic friction simulator: room view (left) and screen snapshot (right).

There are several challenges to implementing these haptic components. Because the Falcon device has a limited range of movement (i.e., physical working volume), it is possible to push the virtual block to an unreachable physical area if enough force is applied. Since the H3D API does not provide any tools for boundary implementation, we impose special boundary conditions on the virtual objects in the scene. We accomplished this by monitoring the block's momentum and position, and defining a range for the block movement in either direction. If the student attempts to push the block beyond these limits, a net force of zero is sent to the block to keep it stationary. If, however, enough force is applied, the block will continue moving according to its momentum and may go beyond the set boundaries. To deal with this case, we invert the block's momentum. The user will interpret this as the cube running into an invisi-ble wall when the boundaries are reached. To simplify user interaction, we also constrained the movement of the block to one axis, movement up and down on the inclined plane.

## 4  Experimental Design and Results

The main goals of the simulator are to enhance student learning, capture student attention, and involve undergraduates in interdisciplinary research. We also want to promote the widespread educational use of this simulator, so we carefully considered the cost factor. A detailed discussion regarding cost will be presented in the conclusion section. In what follows, we describe the experimental framework used to objectively and subjectively measure the simulator's efficiency.

### 4.1  Simulator Efficiency Assessment

In the spring and summer of 2011, we performed several sets of experiments to determine the impact of the simulator in an introductory college level Physics course. We had a total of 86 participants in the experiments.

Before participating in the learning activity, the students took a pre-test, which aimed to evaluate their prior knowledge for learning the subject unit. The pre-test showed that most students had only a rudimentary knowledge of static and kinetic friction, with the average score being 36.7% (random chance would yield a score of 19.7%).

After the pre-test, the students received a 50-minute conventional lecture about static and kinetic friction. The lecture was followed by a post-test about static and kinetic friction.

Post-test results were used to divide the students into two groups (A and B, illustrated in Fig.5) such that each group had equivalent post-test performance. A t-test on the post-test scores of the two groups showed no significant difference ($t=1.49$, $p>.05$), implying that the groups had equivalent theoretical knowledge before participating in the laboratory activity.

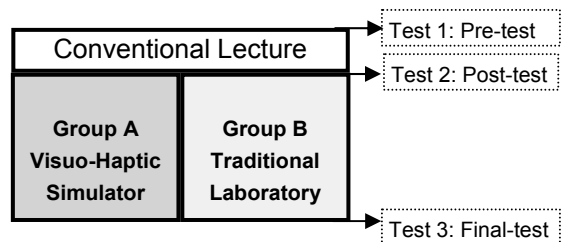

**Fig. 5.** Assessment – groups and tests.

After the division into groups, group A performed lab experiments using the visuo-haptic simulator while students in group B performed similar experiments in a traditional laboratory setup (see Fig. 2).

Both the traditional physics laboratory and the visuo-haptic lab had a paper laboratory handout which provided the students with explanations on how to set up and interact with the blocks on an inclined plane.

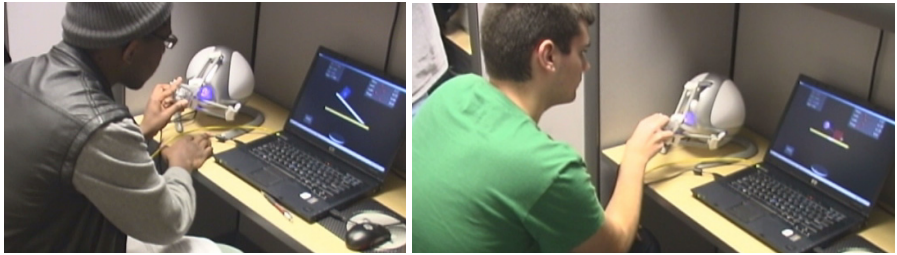

**Fig. 6.** Students in group A, experimenting with the simulator.

Each student had 15 minutes of hands-on work with the simulation (as illustrated in Fig. 6) and 15 minutes of observation. A final test was administered to all students at the same time. The final test results are provided in Fig. 7.

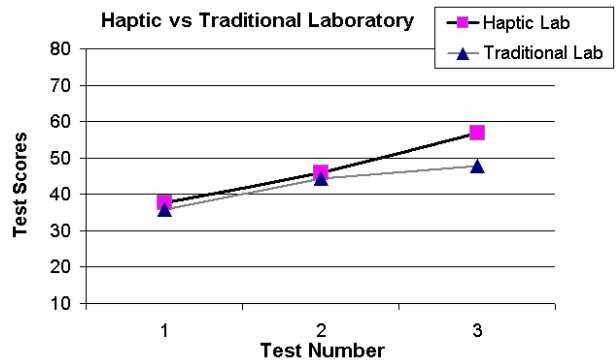

**Fig. 7.** The average test scores for group A and B.

For each group, the normalized gain between the second and third tests was calculated as (Test 3 − Test 2)/(100-Test 2). This metric, therefore, provides the gain as a fraction of the maximum possible gain that could have been achieved between the two tests.

When averaged across all participants, the normalized gain for students using the haptic simulation was 0.182. For students in the traditional physics group, the normalized gain was actually slightly negative at -0.011. Through the use of a t-test, we determined that the chance of an outcome like this occurring if the null hypothesis were correct is 1.2%.

## 4.2 Attitude Surveys and Student Attention Stimulation

An important side-effect of the simulator is student attention. The unfamiliarity with the haptic user interface stirs the students' curiosity and stimulates their attention.

In spring and summer of 2011, we performed an attitude survey with group A, the one involved in the haptic simulation. To better understand the students' perception of the use of the haptic learning system, this study also collected the students' feedback in terms of perceived *usefulness* and perceived *usability* (i.e. ease of use and learnability) of the simulator.

The physics students interviewed enjoyed the simulator's capability to provide novel perspectives on friction. Most students agreed that the simulator effectively demonstrated both static and kinetic friction from a novel perspective.

The attitude surveys were very helpful in improving the simulator's user interface. From the survey, we concluded that navigating the 3D environment was the main problem students had. At first, many students had trouble aligning the Falcon's virtual pointer with the side of the block in order to push it up or down the plane. Some of the students suggested that the color of the pointer should change when it comes in contact with the block, providing additional visual cues in parallel with the haptic ones.

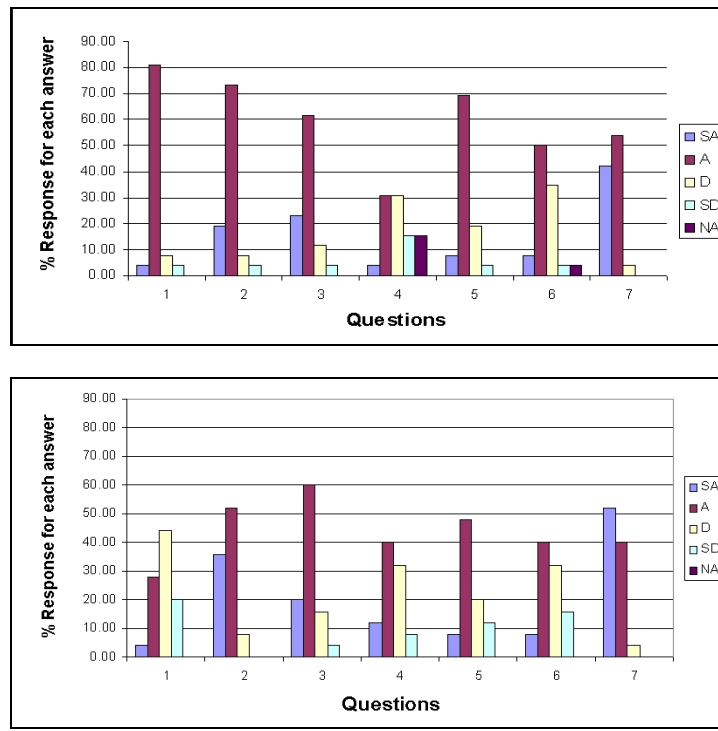

**Fig. 8.** Attitude surveys from Spring (top) and Summer (bottom)

A copy of the attitude surveys is available in the Appendix. While these surveys are heavily subjective, we observed an increase in the students' interest in haptics.

### 4.3 Interdisciplinary Research and Development

Students in Computer Science and Physics were involved in the project from the design to the implementation and testing stages. Since the specific focus of our physics program is Applied Physics and the department has only four full-time physics professors, opportunities for interdisciplinary research efforts with allied departments (e.g., computer science) are especially valuable for the students. Involvement in this project provided the students with a deeper understanding of the illustrated concepts. One of the primary areas of instructional focus in our program is the growing importance of interaction between the physical and virtual worlds. Collection, processing, and analysis of data are the key components of this project as well as modern Applied Physics in general.

## 5 Conclusions

The current cohort of students, known as Millennials or Generation Y, has grown up around technology that is far more sophisticated and abundant than that of their predecessors. Expectations of multimedia entertainment coupled with low attention spans increase the challenge of engaging a student in learning activities by conventional pedagogical methods. Cognitive studies have shown that students are more apt to learn when engaged by the method of exposure. If students could apply their familiarity with modern technology to their learning objectives, they could more easily understand abstract and/or difficult concepts to better relate new information to what they already understand.

Research in psychology demonstrates that learning styles vary from student to student, and that students have diverse learning needs depending on their cognitive styles and abilities. Various brain regions involved in spatial tasks are activated by the synthesis of multiple sensory inputs. Kinesthetic learners make up about 15% of the population and struggle to learn by just reading or listening [18]. We strongly believe that the application of haptic technology to enhance learning of difficult or abstract concepts in science will improve not only the student's laboratory experience, but also a student's attention and retention in the field. Visuo-haptic applications can improve student learning if simulations are carefully chosen by interdisciplinary teams. Moreover, haptics may provide a medium to learn by doing, through first-person experience.

### 5.1 The Cost Factor

The cost of haptic devices is now significantly lower than a few years ago, which makes them affordable augmentations to existing science laboratories. Since the majority of these laboratories are already equipped with computers, the addition of a haptic hardware interface is often as trivial as installing a mouse would be. We chose the Novint Falcon due to its low cost and device characteristics (small working volume and maximum force values) sufficient for simulating friction. In terms of hardware for visualization, one solution is inexpensive 3D red-and-blue or polarized glasses. The software components are also attainable due to their low cost. The X3D standard, supported by several plug-ins, allows for rapid development of 3D virtual scenes in a Web browser and keeps the graphical user interface and 3D environment navigation intuitive since most students are already experts at Web browsing. The H3D API developed by SenseGraphics is a freely available library for developing haptic applications and is closely related to the X3D standard, thus providing interoperability.

### 5.2 Goals

It is important to remember that our goal is not the replacement of traditional learning tools that work well. We explore concepts and paradigms for which a visuo-haptic simulation will enable a better understanding. We envision such environments augmenting rather than replacing existing teaching methods. We are strongly convinced that there are many abstract ideas in science which cannot be cheaply or easily realized physically in a pedagogically useful manner, but that would be well-illustrated through the visuo-haptic approach.

An efficient learning environment must provide excellent perceptual integration, which is not only task-dependent, but might be even more difficult to attain than the technical integration. Discovering and defining simulators and training tools that would benefit from the haptic feedback are also challenging tasks. One must identify the concepts that lend themselves best to such simulation, and then design a learning experience rather than merely a simulation.

While the technical integration of the haptic sensation is important, so is the measurement of its impact on learning. After two years of experimentation we have identified several problems in the practical and objective assessment of the simulator. A balanced group composition in terms of test scores is required, and sufficient warm-up trials with the haptic devices are also necessary.

We experienced several setup issues (e.g., time constraints were important as the haptic devices had to be attached to laptops and the applications preconfigured to be ready during class). Scheduling was complicated by the demands on student and instructor time, as well as the large number of other classes using the physics classrooms. Regardless of the technical integration issues, we have proven that carefully designed and deployed haptic simulators can have a positive role in physics education.

**Appendix: Attitude Survey**

Check one:

- Freshman
- Sophomore
- Junior
- Senior

For each question, please select one of the following:

- SA    - Strongly Agree,
- A     - Agree,
- D     - Disagree,
- SD    - Strongly Disagree,
- NA    - Not Applicable.

1. The Novint Falcon haptic device was easy to use.

   SA    A    D    SD    NA

2. The simulator was effective in demonstrating the behaviour of static friction.

   SA    A    D    SD    NA

3. The simulator was effective in demonstrating the behaviour of kinetic friction.

   SA    A    D    SD    NA

4. The simulator was more effective in illustrating friction than a conventional laboratory experiment.

   SA    A    D    SD    NA

5. The environment was intuitive and easy to understand.

   SA    A    D    SD    NA

6. It was easy to navigate in the environment.

   SA    A    D    SD    NA

7. It was easy to adjust the parameters that affect the force of friction.

   SA    A    D    SD    NA